\RequirePackage{ifpdf}
\ifpdf
\documentclass{pdftex}
\else
\documentclass{aastex}
\fi

\usepackage{graphics,graphicx}

\shorttitle{Accretion Shock in TW Hya}
\shortauthors{Brickhouse \etal}

\slugcomment{Published online in the Astrophysical Journal Letters,
(ApJ, 760, L21)}

\begin{document}

\def\longrefsmall#1 {\par{\hangindent=10pt \hangafter=1 #1 \par}}
\def\la{\ifmmode\stackrel{<}{_{\sim}}\else$\stackrel{<}{_{\sim}}$\fi} 
\def\ga{\ifmmode\stackrel{>}{_{\sim}}\else$\stackrel{>}{_{\sim}}$\fi} 
\newcommand{\longref}[1]{\par\noindent{\hangindent=20pt \hangafter=1 #1 \par}}
\newcommand{\logte}{$\log\,T_e[{\rm K}]$}
\newcommand{\etal}{{et~al.}}
\newcommand{\chandra}{{\it Chandra}}
\newcommand{\xmm}{{\sl XMM-Newton}}
\newcommand{\Chandra}{{\it Chandra}}
\newcommand{\XMM}{{\sl XMM-Newton}}
\newcommand{\Sherpa}{{\it Sherpa}}
\newcommand{\sherpa}{{\it Sherpa}}
\newcommand{\eg}{{\it e.g.}}
\newcommand{\ie}{{\it i.e.}}
\newcommand{\NH}{\mbox{$N_{\rm H}$}}        

\title{X-Ray Determination of the Variable Rate of Mass Accretion onto TW Hydrae}

\author{N. S. Brickhouse, S. R. Cranmer, A. K. Dupree, 
  H. M. G\"{u}nther, G. J. M. Luna\footnote{Current address: Instituto de Astronomia y Fisica del Espacio,
(IAFE), Buenos Aires, Argentina},  and S. J. Wolk}
\affil{Harvard-Smithsonian Center for Astrophysics, 60 Garden Street, Cambridge, MA 02138}

\begin{abstract}

Diagnostics of electron temperature ($T_e$), electron density ($n_e$),
and hydrogen column density ($N_H$) from the \chandra\ High Energy
Transmission Grating spectrum of He-like \ion{Ne}{9} in TW Hydrae (TW~Hya), in
conjunction with a classical accretion model, allow us to infer the
accretion rate onto the star directly from measurements of the
accreting material. The new method introduces the use of the absorption of
\ion{Ne}{9} lines as a measure of the
column density of the intervening, accreting material. On average, the derived mass
accretion rate for TW~Hya is $1.5 \times 10^{-9}$ M$_{\odot}$ yr$^{-1}$, for a
stellar magnetic field strength of 600 Gauss and a filling factor of
3.5\%. Three individual \chandra\ exposures show statistically
significant differences in the \ion{Ne}{9} line ratios, indicating changes in
$N_H$, $T_e$, and $n_e$ by factors of 0.28, 1.6, and 1.3,
respectively.  In exposures separated by 2.7 days, the observations
reported here
suggest a five-fold reduction in the accretion rate. This powerful
new technique promises to substantially improve our understanding of
the accretion process in young stars.

\end{abstract}


\section{Introduction}

In the standard scenario for magnetic accretion onto classical T Tauri
stars (CTTS), material accelerates along magnetic field lines from
a truncated accretion disk toward the star, producing an X-ray
emitting shock of a few MK near the stellar surface. Many
accretion-related phenomena observed in ultraviolet and optical
spectra, e.g. excess ultraviolet continuum and the filling in or
``veiling'' of photospheric absorption lines by hot continuum, are
interpreted as the stellar atmosphere's response to the 
X-ray shock. Optical, near infrared, and ultraviolet emission lines also show
accretion-related broadening of a few hundred km~s$^{-1}$. Optical
spectral signatures facilitate a number of indirect techniques for
estimating the mass accretion rate and hot spot filling factor.  High
resolution X-ray spectra of CTTS with the \chandra\ and \xmm\ gratings
provide line ratio diagnostics of the electron temperature ($T_e$) and
electron density ($n_e$) in the shock which supports this basic magnetospheric
accretion model. Diagnostics of the X-ray shock in principle can
provide a more direct measure of the accretion parameters
(e.g. G\"{u}nther 2011), but they have often significantly underestimated the
accretion rate relative to more traditional methods (e.g. Calvet \&
Gullbring 1998; Hartigan \& Kenyon 2003).

Kastner et~al. (2002) presented the first evidence for high $n_e$ in a
CTTS using the density-sensitive He-like \ion{Ne}{9} diagnostic ratios
from a \chandra\ High Energy Transmission Grating (HETG) spectrum of
TW~Hydrae (TW~Hya). This high density in TW~Hya was also observed
using the \xmm\ Reflection Grating Spectrometer (Stelzer~\& Schmitt
2004). High $n_e$ at relatively low $T_e$ has now been measured in a
number of CTTS systems (G\"{u}del~\& Naz\'{e} 2010, and references
therein).

TW~Hya is nearby (57~pc) and, with an inclination angle of
7$^{\rm{o}}$ (Qi et~al. 2004), provides a pole-on view of the
star where accretion occurs. \chandra\ observed TW~Hya with the HETG for 489~ks in 2007 to
obtain a spectrum with high signal-to-noise ratio.  The spectrum
provides a number of diagnostic line ratios not yet obtained for
other CTTS systems (Brickhouse et~al. 2010, hereafter Paper~I). In
particular, diagnostics indicate three emission components: a hot
corona ($\sim$10~MK), an accretion shock (2.5~MK) in good agreement
with model predictions, and a large post-shock region ($<$2~MK) with 30
times the mass of the shock front. \ion{O}{7} line ratio diagnostics
of this third component rule out its origin in the radiative cooling
zone of the shock and suggest instead its production by
accretion-driven heating of surrounding stellar atmosphere.

In this Letter we present new results from the 2007 \chandra\ spectra
of TW~Hya, focusing on the accretion shock component. Line ratio
diagnostics measured from three different pointings show
significant variation.  \ion{Ne}{9} line ratio diagnostics constrain
the fundamental parameters of a basic accretion model. In particular,
the absorbing column density $N_H$ derived from \ion{Ne}{9} line
ratios provides a new diagnostic of the accretion shock.

\section{Spectral Analysis}

\Chandra\ HETG spectra of TW~Hya are analyzed from three different
pointings spanning 2007 February to 2007 March --- observation identifications
(obsids) 7435 (153.3~ks), 7437 (157.0~ks), and 7436
(158.4~ks). A fourth observation of  TW~Hya (obsid 7438) is too short
to provide sufficient signal-to-noise ratio for our purposes here.

Table~1 presents fluxes from the three pointings for the strongest
observed lines of \ion{Mg}{11}, \ion{Ne}{10}, \ion{Ne}{9}, \ion{O}{8},
and \ion{Fe}{17}.  In Paper~I we argued that these ions are formed in
the cooling column of the shock, whereas lines
from \ion{Mg}{12}, \ion{Si}{13}, and \ion{Si}{14} are formed in a
stellar corona and lines from \ion{O}{7} are
produced in a large post-shock region associated with the soft
excess found in other CTTS (G\"{u}del~\& Telleschi 2007; Robrade~\&
Schmitt 2007). The \ion{Ne}{10} lines are likely to be formed in both the accretion shock
and the hot corona. We include them here in order to assess this
assumption.

The He-like \ion{Ne}{9} system provides line ratio diagnostics for
$T_e$, $n_e$, and $N_H$ (Paper~I). Fortunately, the atomic data for
this ion are now accurate for diagnostic studies (Chen et~al. 2006).
Figure~1 shows the theoretical $T_e$-dependent curve for \ion{Ne}{9}
(Foster et~al. 2012) with the measured G-ratios [(forbidden plus intercombination)/resonance] 
overplotted. During the third observation $T_e$ is significantly
higher than during the first two observations. This change in $T_e$
provides one of the primary motivations for this study.

The observed \ion{Ne}{9} series ratios rule out resonance scattering,
consistent with the measured line broadening (Paper I). The line
optical depth is reduced by two orders of magnitude due to the modeled
filling factor (Table~2) and turbulent broadening.  Figure~2 shows
$N_H$ inferred from the measured \ion{Ne}{9} resonance He$\alpha$
($\lambda$13.45) /He$\beta$ ($\lambda$11.54) ratios, using a neutral
to near-neutral photoelectric absorber.  $N_H$ is smaller in the third
observation compared with the first two, providing additional
motivation for this study. To derive $N_H$ from the
He$\alpha$/He$\beta$ line ratio, the $T_e$-dependence is first
determined from the observed G-ratio.  Theoretical curves of
He$\alpha$/He$\beta$ are then plotted as functions of $N_H$, for
different values of $T_e$ ranging from 1.6 to 4.0 MK.  Table~2 gives the derived
values for $T_e$, $N_H$, and $n_e$. We note that
$n_e$ derived from the R-ratio (forbidden/intercombination; not shown)
is only marginally significantly different in the third
pointing. Since the values of $n_e$ from all three pointings are
more than an order of magnitude higher than observed from purely
coronal sources at \ion{Ne}{9} temperatures (Sanz-Forcada
et~al. 2003), the assumption that \ion{Ne}{9} forms primarily in the
accretion shock remains justified.

\ion{Fe}{17} also originates primarily in the shock (Paper~I). We use
the observed $\lambda$15.01 (``3C'') to $\lambda$15.26 (``3D'') line
ratios to obtain $T_e \approx$ 3.0, 2.1, and 4.6 ~MK for pointings 1,
2, and 3, respectively. The temperature sensitivity of 3C/3D arises
from an inner shell \ion{Fe}{16} line blend with 3D (Brown
et~al. 2001). While these $T_e$ values are not entirely consistent
with those from \ion{Ne}{9}, they support the trend for higher $T_e$
during the third pointing.  Notably, the ratio for the third
observation is consistent with the absence of \ion{Fe}{16}, as
expected at the higher temperature.

We can test the overall consistency of the derived parameters by
comparing the observed line fluxes with predictions. For a single
component model, the observed line intensity $I_{obs} = W_\lambda~ A~
\varepsilon(T_e)~ EM/(4 \pi R^2)$, where $W_\lambda$ is the
wavelength-dependent transmission through the absorber (Morrison~\&
McCammon 1983; Wilms et al. 2000), $A$ is the relative elemental abundance, $\varepsilon$
is the emissivity from AtomDB (Smith et~al. 2001; Foster~et al. 2012),
$EM$ is the emission measure ($EM = n_e^2 V$, where $V$ is the volume)
and $R$ is the distance to the source. 
The absorption model assumes solar
abundances, which could introduce an error as large as a factor of two
in $N_H$.

We use the summed observation to determine $A$ and then
predict fluxes for the three pointings. Abundance ratios
for oxygen, magnesium, and iron compared to neon are in the range of relative abundances found in
Paper~I. The observed fluxes of \ion{Ne}{9} $\lambda$13.45 along with
the values of $n_e$, $T_e$, and $W_\lambda$ determined from the
\ion{Ne}{9} diagnostics, give $V$ for each pointing.  The predicted
fluxes of the resonance lines \ion{Mg}{11} $\lambda$9.17, \ion{O}{8}
$\lambda$18.97 and \ion{Fe}{17} $\lambda$15.01 are reasonably
consistent with observations; however, the observed fluxes show
smaller differences than predicted. For
all three ions, a smaller difference in the transmission
$W_\lambda$ gives better agreement. For example, decreasing
$N_H$ from $\sim 3.0 \times 10^{21}$ cm$^{-2}$ to $\sim 2.2 \times
10^{21}$ cm$^{-2}$ brings the \ion{O}{8} line for the first two pointings
into good agreement, while increasing
$N_H$ from 0.9 $\times 10^{21}$ cm$^{-2}$ to a value between 1.0 and
1.3 $\times 10^{21}$ cm$^{-2}$ brings the line fluxes for all three
ions into excellent agreement for the third pointing. The consistency
for \ion{Mg}{11} and \ion{Fe}{17} is somewhat worse for the first two
pointings, probably because their emissivity functions are steeply
dropping at such low $T_e$. Given order of magnitude differences
in emissivities and transmission over the range in $T_e$ and
$N_H$, the good agreement among the different shock
diagnostics is remarkable.

We predict \ion{Ne}{10} Ly$\alpha$ in a similar manner. Our
predictions are low by a factor of ten for the first two
pointings, but by only a factor of two for the third. \ion{Ne}{10} arises
primarily from the corona, but has a larger accretion contribution
during the third pointing compared with the other two.

\section{Determining the Parameters of the Accretion Shock}

The observed variation of $N_H$ provides new support for the idea that
the accreting pre-shock stream is the absorber of the shocked, X-ray
emission (e.g. Lamzin 1999; Gregory et~al. 2007; Paper~I),
since other potential sources of absorption (e.g. interstellar,
circumstellar) are not expected to vary significantly. Moreover, the
nearly face-on view of the TW~Hya system makes the disk itself an implausible
absorber. To explore what the pre-shock gas as absorber implies about
the geometry of the system, we use the strong shock condition to
obtain the electron density of the pre-shock gas ($n_e$/4) from the observed density $n_e$ of the
shock. We assume the observed column density $N_H$ is through the same
pre-shock gas, and estimate the path
length $<l> = 4 N_H/n_e$. From \ion{Ne}{9},\footnote{These values are
from the summed spectrum, with a small correction for the
$T_e$-dependence of the \ion{Ne}{9} R-ratio, following Smith et
al. (2009).}  $N_H=1.8 \times 10^{21}$ cm$^{-2}$ and $n_e=3.2 \times
10^{12}$ cm$^{-3}$, and thus $<l> = 2.2 \times 10^4$ km. This path
length is much smaller than the stellar radius, and thus geometrical
effects due to the curvature of the accretion column are
insignificant.  This estimate of $<l>$ is also an order of
magnitude larger than the length of the post-shock cooling zone
(derived in Paper~I). Absorption by the cooling column appears to be
ruled out. Absorption by intervening stellar atmosphere cannot be
ruled out, although recent hydrodynamic models produce an ``observable'' shock
in the upper chromosphere for conditions similar to those of TW~Hya
(Sacco et~al. 2008; 2010). Thus we proceed with
the assumption that the pre-shock stream is the only absorber.

In the standard model of magnetospheric accretion (Calvet~\&
Gullbring 1998; G\"{u}nther et~al. 2007), the shock temperature can
only vary if the accreting material originates at different distances
from the star.  K\"{o}nigl (1991) gives an expression for the inner
truncation radius of the disk that depends only on the mass accretion
rate $\dot{M}_{\rm acc}$ and the surface poloidal magnetic field
strength $B_{\ast}$, in addition to the stellar mass $M_{\ast}$ and
radius $R_{\ast}$. Thus the observed changes in $T_e$ indicate
variations in $\dot{M}_{\rm acc}$, $B_{\ast}$, or both.

Together $T_e$, $n_e$, and $N_H$ constrain the fundamental parameters
of the standard accretion model, namely $\dot{M}_{\rm acc}$,
$B_{\ast}$, and filling factor $f$. We assume a solar-type magnetic dipole, aligned
with the rotation axis of the star (Cranmer 2008, 2009). Material originates in the
gaseous accretion disk between the inner truncation radius $r_{in}$ and an
outer radius $r_{out}$, both of which are dependent variables in the
model. The accretion stream shocks near the surface of the star,
forming a circumstellar ring at a latitude determined by the dipole
field geometry and $r_{in}$. We have run a cube of $50 \times 50
\times 50$ models, independently varying $\dot{M}_{\rm acc}$,
$B_{\ast}$, and $f$ for comparison with observations. The value of
$\dot{M}_{\rm acc}$ is calculated for the whole star, i.~e. in
the pole-on case the model accretion rate is twice what we
observe. The model parameters are
given in Table~2 and illustrated in Figure~3 for the case where the
free fall velocity is calculated from $r_{in}$. Similar results
are found if we compute the free fall velocity as the square root of
the product of the velocities derived from material released at the
inner and outer edges of the disk, respectively.

We illustrate the model constraints using the diagnostic lines of
\ion{Ne}{9}, since \ion{Ne}{9} is the only He-like ion that produces
lines strong enough to determine $T_e$, $n_e$, and $N_H$ from the
accretion shock in all three pointings.  For the summed
data and for each pointing, approximately 100 models, or
0.1\% of the 125,000 models computed, are found to agree to within
1$\sigma$ of the derived $T_e$, $n_e$, and $N_H$.
The summed spectrum, gives the average values $\dot{M}_{\rm acc} = 1.5 \times
10^{-9}$ M$_{\odot}$ yr$^{-1}$, $B_{\ast} = 600$~G, and $f = 3.5$\%. The
accreting material originates at $< 3R_{\ast}$.

Here we have assumed a single $T_e$ and $n_e$ model for the shock.  In
Paper~I we found $f$ = 1.1\% and
$\dot{M}_{\rm acc}$ = $4 \times 10^{-10}$ M$_{\odot}$ yr$^{-1}$ by
computing $T_e$, $n_e$ and the resulting line emission as functions of
depth in the post-shock cooling
zone, and assuming a truncation radius of 4.5~$R_{\ast}$.
We then
integrated the emission over the cooling zone and compared to
diagnostic line fluxes from 
the summed spectrum. The shock front temperature was
3.4~MK, compared with 2.5~MK here, while the shock-front electron
density was slightly smaller at $ 2.3 \times 10^{12}$ cm$^{-3}$. The
higher accretion rate derived here arises primarily from the three
times larger $f$, with somewhat larger $n_e$ offset by somewhat
smaller $T_e$. 

We compare the derived $\dot{M}_{\rm acc} = 1.5 \times 10^{-9}$
M$_{\odot}$ yr$^{-1}$ to $0.4$, $1.3$, and $5 \times 10^{-9}$ M$_{\odot}$
yr$^{-1}$, from Muzerolle et~al. (2000), Donati et~al. (2011), and
Batalha et~al. (2002), respectively. Curran et~al. (2011) report
consistent rates from optical diagnostics.  Our value agrees well
with these estimates. Donati et~al. (2011) demonstrate using
spectropolarimetric signatures that a strong quadrupole magnetic field
(2.5 to 2.8 kG) dominates the magnetic structure in the photosphere of TW~Hya; they also
report a weaker dipole component (400 to 700 kG) which is capable of
disrupting the accretion disk at 3 to 4 $R_{\ast}$, also consistent
with our results. Several authors report significantly lower mass
accretion rates from X-ray observations: $1.8 \times 10^{-10}$ (Curran
et~al. 2011); $2 \times 10^{-11}$ (Stelzer~\& Schmitt 2004); $2 \times
10^{-10}$ (G\"{u}nther et~al. 2007); and, $6$ to $7 \times 10^{-11}$
(G\"{u}nther 2011) M$_{\odot}$ yr$^{-1}$. All of these studies use a
lower $N_H$ value determined from global fits to the spectra,
accounting for a factor of three or larger difference between their
accretion rates and ours. The remaining differences may be due to
differences in absolute abundances, filling factors, or relative
accretion and coronal contributions to the spectrum, and to
different spectra and methods. 

The accretion parameters derived from the third  pointing are
strikingly different from those derived from the first two
pointings. $\dot{M}_{\rm acc}$ drops by a factor of 5 while $f$ is
smaller by a factor of 7. The truncation radius moves farther out
by about 70\%. While the derived $r_{in}$ values may be
underestimated by the single component shock, $r_{in}$
cannot be too large or the shock temperature would not show any
measurable variation. Accounting for the systematic bias introduced by
the single component simplification suggests an average $r_{in}$ value
of 4~$R_{\ast}$ from the summed spectrum, with a range from 3.4 to 6.1~$R_{\ast}$ for the three pointings.

The large variations in $\dot{M}_{acc}$ and $f$ are unlikely to be the
result of rotation or viewing angle, and thus it appears that the
accretion rate changes substantially while the stellar magnetic field
changes only slightly, if at all. The single spectrum obtained with
the MIKE double-echelle spectrograph on Magellan during the
third \chandra\ pointing (Dupree et~al. 2012) shows a decrease in
the average blue veiling to $r=0.8$ from earlier values of $r>1.2$
during the second \chandra\ pointing, consistent with a decrease in the
accretion rate. The values of
$\dot{M}_{\rm acc}$ from the three
pointings, 2.8, 2.2, and $0.53 \times 10^{-9}$ M$_{\odot}$
yr$^{-1}$, are all well within the literature values, suggesting that at
least some of the differences in the literature are due to 
intrinsic variability rather than to differences in measurement
method. The observations indicate that at lower accretion rates, the stellar
magnetic field exerts stronger control of the gas, producing a smaller
hot spot at higher latitude, as expected. 

Episodic events at lower latitudes occur in the data of Donati et
al. (2011), but on time-scales much shorter than the time-resolution
of the X-ray spectroscopy. Veiling and spectral line
diagnostics for both accretion and wind showed variability during ground-based
observations contemporaneous with the \Chandra\ observations (Dupree et
al. 2012). The time resolution of these variation signatures is of the
order of several minutes, again much shorter than the effective time
resolution of the X-ray spectroscopy. X-ray light curves
derived from accretion lines show variability on timescales of a few
ks (see also Paper~I), although these variations are not correlated
with the optical photometry, and  an enhanced accretion event (during
pointing 2) is followed by systematic changes in the optical emission
lines (Dupree et~al. 2012).
A decrease in the accretion rate leads to a complex response by the X-ray light
curve: a decrease in intrinsic emission measure is countered by an
increase in observed flux due to a decrease in the soft X-ray
absorption by the accreting column itself.

\section{Conclusions}

Significant variability in X-ray accretion diagnostics occurs during the
\Chandra\ HETG observations of TW~Hya. In particular, variability of the
column density $N_H$ implicates the pre-shock accreting gas as the
X-ray absorber. $N_H$ and $n_e$ measurements from He-like \ion{Ne}{9}
lines support an absorbing path length short compared to the stellar
radius but long compared to the shock cooling depth. With $T_e$, $n_e$, and $N_H$ together the X-ray data
constrain all the fundamental parameters of the standard accretion
model, namely $\dot{M}_{\rm acc}$, $B_{\ast}$, and $f$.

The resulting average mass accretion rate during the \Chandra\
observation, $\dot{M}_{\rm acc} = 1.5 \times 10^{-9}$ M$_{\odot}$
yr$^{-1}$, based on \ion{Ne}{9} diagnostics alone, agrees
well with optical and ultraviolet techniques. The
observations indicate that the mass accretion rate decreases by a factor
of 5 during one pointing of $\sim$150~ks, compared with two
others. This accretion rate
variation appears to arise from changes in the inner radius of the accretion disk.

\acknowledgments

We acknowledge support from NASA to the Smithsonian Astrophysical
Observatory (SAO) under \Chandra\ GO7-8018X for GJML. NSB and SJW
were supported by NASA contract NAS8-03060 to SAO for the Chandra
X-ray Center. SRC's contribution to this work was supported by 
NASA grant NNG04GE77G to SAO. Support for HMG's contribution was
provided through HST grant GO-12315.01 from NASA.

\clearpage

\begin{deluxetable}{lrrrr}
\tabletypesize{\footnotesize}
\tablewidth{0pc}
\tablecaption{Emission Line Fluxes}
\tablehead{
\colhead{Ion} & \colhead{$\lambda_{ref}$\tablenotemark{a}} &
\multicolumn{3}{c}{Flux (uncorrected for absorption)} \\
 & \colhead{(\AA)} &
\multicolumn{3}{c}{(10$^{-6}$ ph cm$^{-2}$ s$^{-1}$)} 
} 
\startdata
 & & Pointing 1\tablenotemark{b} & Pointing 2\tablenotemark{b} & Pointing 3\tablenotemark{b} \\
\ion{Mg}{11} & 9.17    & 2.5   $\pm$ 0.4 & 1.7   $\pm$ 0.5 & 2.4   $\pm$ 0.4 \\
\ion{Ne}{10} & 9.71    & 1.8   $\pm$ 0.8 & 3.1   $\pm$ 0.5 & 2.8   $\pm$ 0.6 \\
\ion{Ne}{10} & 10.24   & 7.0   $\pm$ 0.7 & 9.0   $\pm$ 2.0 & 8.7   $\pm$ 0.8 \\
\ion{Ne}{9}  & 11.00   & 7.3   $\pm$ 0.7 & 6.6   $\pm$ 1.0 & 7.9   $\pm$ 1.0 \\
\ion{Ne}{9}  & 11.54   & 19.9  $\pm$ 1.5 & 23.7  $\pm$ 1.8 & 24.2  $\pm$ 1.6 \\
\ion{Ne}{10} & 12.13   & 63.8  $\pm$ 3.1 & 74.1  $\pm$ 3.3 & 77.2  $\pm$ 3.4 \\
\ion{Fe}{17} & 12.27   & 3.2   $\pm$ 0.8 & 4.7   $\pm$ 1.0 & 2.1   $\pm$ 0.6 \\
*\ion{Ne}{9}  & 13.45   & 128.3 $\pm$ 6.9 & 144.7 $\pm$ 7.0 & 187.8 $\pm$ 8.0 \\
\ion{Ne}{9}  & 13.55   & 93.5  $\pm$ 5.5 & 97.8  $\pm$ 5.9 & 115.4 $\pm$ 6.2 \\
\ion{Ne}{9}  & 13.70   & 47.4  $\pm$ 4.0 & 50.4  $\pm$ 3.4 & 53.9  $\pm$ 4.3 \\
\ion{Fe}{17} & 15.01   & 28.6  $\pm$ 3.5 & 29.2  $\pm$ 3.5 & 43.4  $\pm$ 4.1 \\
\ion{Fe}{17} & 15.26   & 11.7  $\pm$ 2.5 & 15.0  $\pm$ 3.2 & 15.8  $\pm$ 2.4 \\
\ion{O}{8}   & 16.01   & 27.1  $\pm$ 3.2 & 24.3  $\pm$ 3.8 & 32.7  $\pm$ 3.9 \\
\ion{Fe}{17} & 16.78   & 25.6  $\pm$ 4.7 & 22.4  $\pm$ 4.0 & 25.3  $\pm$ 4.3 \\
\ion{Fe}{17} & 17.05   & 25.5  $\pm$ 7.6 & 28.4  $\pm$ 4.5 & 31.8  $\pm$ 4.3 \\
\ion{Fe}{17} & 17.10   & 23.0  $\pm$ 6.4 & 28.7  $\pm$ 4.6 & 29.4  $\pm$ 4.2 \\
\ion{O}{8}   & 18.97   & 201.3 $\pm$ 15. & 210.1 $\pm$ 15. & 201.0 $\pm$ 16. \\

\enddata
\tablenotetext{a}{Reference wavelengths from AtomDB v2.0 (Foster et
  al. 2012).
  Multiplets are intensity-weighted averages.}  
\tablenotetext{b}{Pointings 1, 2, and 3 correspond to mid times 48.0,
  58.5, and 61.2 (JD-2454100), respectively.}
\end{deluxetable}

\clearpage
\begin{deluxetable}{lcccc}
\tabletypesize{\footnotesize}
\tablewidth{0pc}
\tablecaption{Physical Conditions and Accretion Model Parameters}
\tablehead{
\colhead{Parameter} &
\colhead{Pointing 1\tablenotemark{a}} &
\colhead{Pointing 2\tablenotemark{a}} &
\colhead{Pointing 3\tablenotemark{a}} &
\colhead{Total\tablenotemark{b}}
}
\startdata 
$T_e$ (MK)\tablenotemark{c}                  & 1.9$^{+0.4}_{-0.3}$                             & 2.3$^{+0.4}_{-0.3}$      & 3.1$^{+0.7}_{-0.4}$       &2.5$^{+0.3}_{-0.2}$       \\
$n_e$ (10$^{12}$ cm$^{-3}$)\tablenotemark{c} & 3.0$^{+0.4}_{-0.3}$                                 & 3.1$^{+0.3}_{-0.3}$       & 3.9$^{+0.3}_{-0.4}$       & 3.2$^{+0.2}_{-0.2}$        \\
$N_H$ (10$^{21}$ cm$^{-2}$)\tablenotemark{d} & 3.2$^{+1.1}_{-0.8}$                          & 2.8$^{+0.8}_{-0.8}$ & 0.9$^{+0.8}_{-0.5}$  & 1.8$^{+0.6}_{-0.4}$  \\
$\dot{M}_{\rm acc}$ (10$^{-9}$ M$_{\odot}$ yr$^{-1}$)\tablenotemark{e} & 2.78 (2.02--3.56)  & 2.19 (1.68--2.95)   & 0.53 (0.19--1.15)    & 1.45 (1.05--1.84) \\
Poloidal $B_{\ast}$ ($G$)\tablenotemark{e} & 523. (389.--679.)                           & 628. (456.--797.)   & 770. (359.--1510.)   & 614. (494.--797.) \\
Filling Factor (\%)\tablenotemark{e}        & 8.19 (5.08--12.3)                             & 5.69 (3.91--8.63)   & 0.95 (0.29--2.46)    & 3.46 (2.52--4.59)  \\  
r$_{in}$/R$_{\ast}$\tablenotemark{e}  & 1.77 (1.58--1.99)                                  & 2.11 (1.84--2.42)   & 3.60 (2.58--4.89)    & 2.34 (2.07--2.60) \\
r$_{out}$/R$_{\ast}$\tablenotemark{e} & 2.23 (1.86--2.67)                                  & 2.57 (2.10--3.04)   & 3.82 (2.67--5.33)    & 2.68 (2.33--3.14) \\
Latitude\tablenotemark{f}             & 44.6$\arcdeg$ & 49.0$\arcdeg$ & 58.7$\arcdeg$ & 50.8$\arcdeg$ \\
\enddata
\tablenotetext{a}{Pointings 1, 2, and 3 correspond to mid times 48.0,
  58.5, and 61.2  (JD-2454100), respectively.}
\tablenotetext{b}{Derived properties $T_e$, $n_e$ and $N_H$ are from Brickhouse
et~al. (2010), except for a minor revision to $n_e$, for which we have
taken the $T_e$-dependence of the $n_e$ diagnostic into account here. The error on $N_H$ is estimated in this work.}
\tablenotetext{c}{Errors on $T_e$ and $n_e$ use 1$\sigma$ errors on the observed line ratios.}
\tablenotetext{d}{Errors on $N_H$ include 1$\sigma$ errors on $T_e$ (from the G-ratio) as well as 1$\sigma$ errors on the He$\alpha$/He$\beta$ ratio.}
\tablenotetext{e}{Values given are the average values from the models that
meet the criteria for each time segment, described in the text. The range in parentheses gives the minimum
and maximum values for the same set of models.}
\tablenotetext{f}{Average latitude of the field line which the
accretion stream follows.}
\end{deluxetable}

\clearpage
\begin{figure}
\includegraphics[width=6.0in]{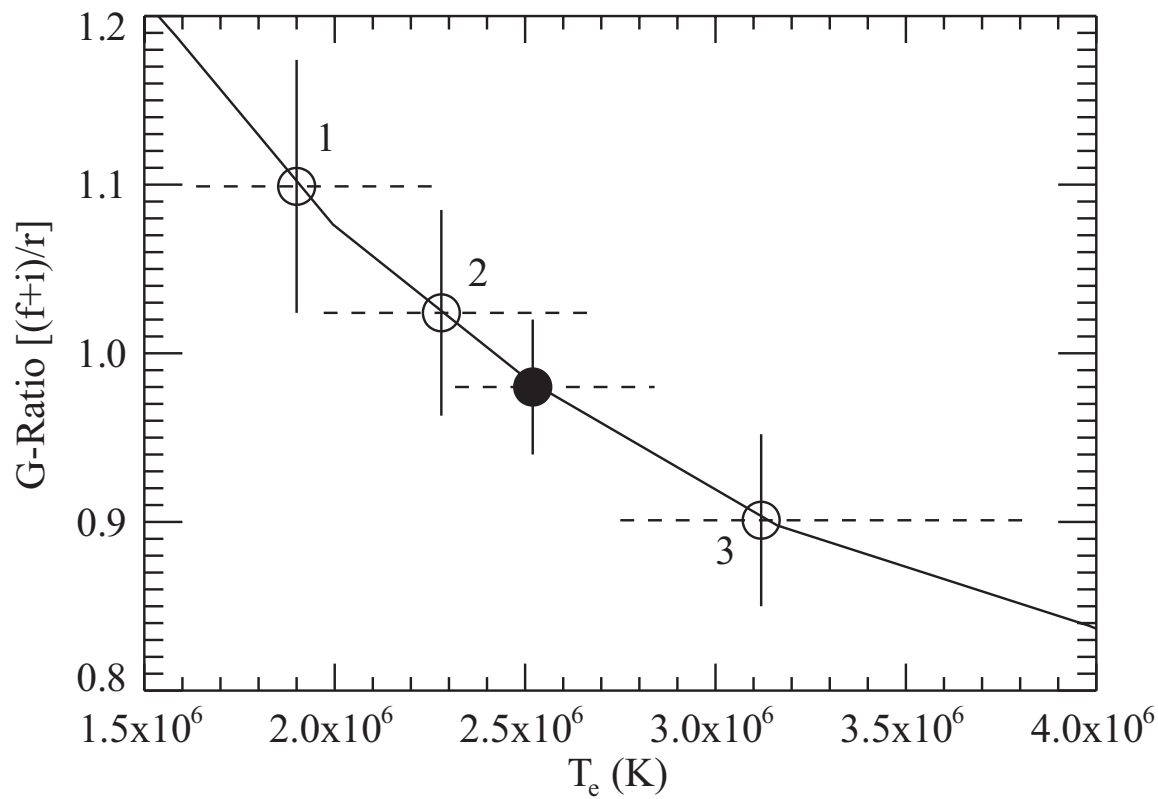}
\caption{Theoretical G-ratio (i.e., ratio of forbidden plus
intercombination to resonance line fluxes) for \ion{Ne}{9} as a
function of $T_e$ (solid curve) from AtomDB (Foster et~al. 2012; Chen
et~al. 2006). Overplotted
are the observed ratios with 1$\sigma$ errors (vertical error bars) for the individual
observations (open circles, numbered in time order) and for the summed spectrum (solid
circle). Horizontal error bars (dashed lines) show how the
errors on the observed ratios translate to the errors on $T_e$.
}
\end{figure}

\clearpage
\begin{figure}
\includegraphics[width=6.0in]{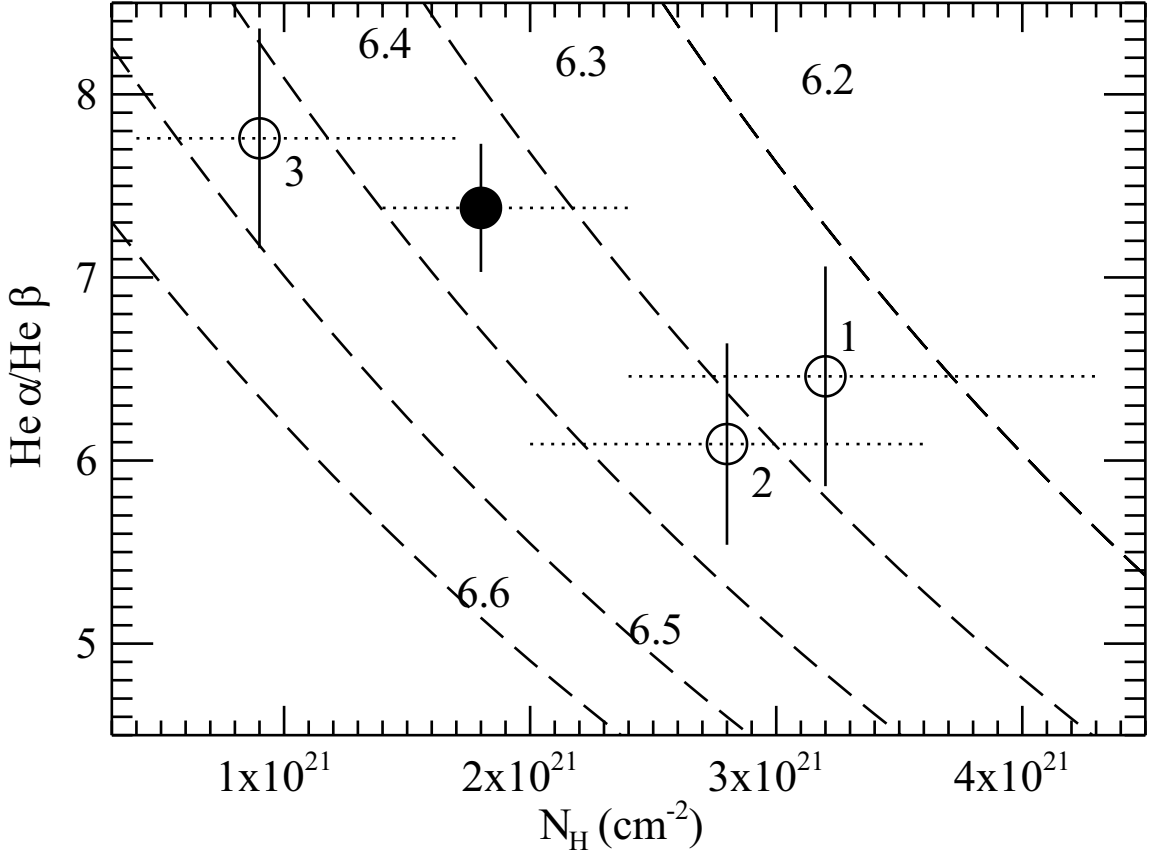}
\caption{Ratio of the flux of the resonance line (He$\alpha$) to that
  of He$\beta$ as a function of hydrogen column density (dashed
  curves), for given log~[$T_e$]~(K) from 6.2 to 6.6 as
  labeled. Absorption models are from Morrison \& McCammon (1983) and
  line emissivities are from Foster et~al. (2012). Overplotted are the observered ratios with 1$\sigma$ errors
  (vertical error bars) for the individual observations (open circles,
  numbered in time order)
  and for the summed spectrum (solid circle). Observed points are
  placed on the curve appropriate for their $T_e$ as determined from
  their G-ratios. Horizontal error bars (dotted lines) show how the
  errors both on the observed He$\alpha$/He$\beta$ ratios and on $T_e$
  (determined from the G-ratios) translate to the errors on $N_H$.
}
\end{figure}

\clearpage
\begin{figure}
\includegraphics[width=6.0in]{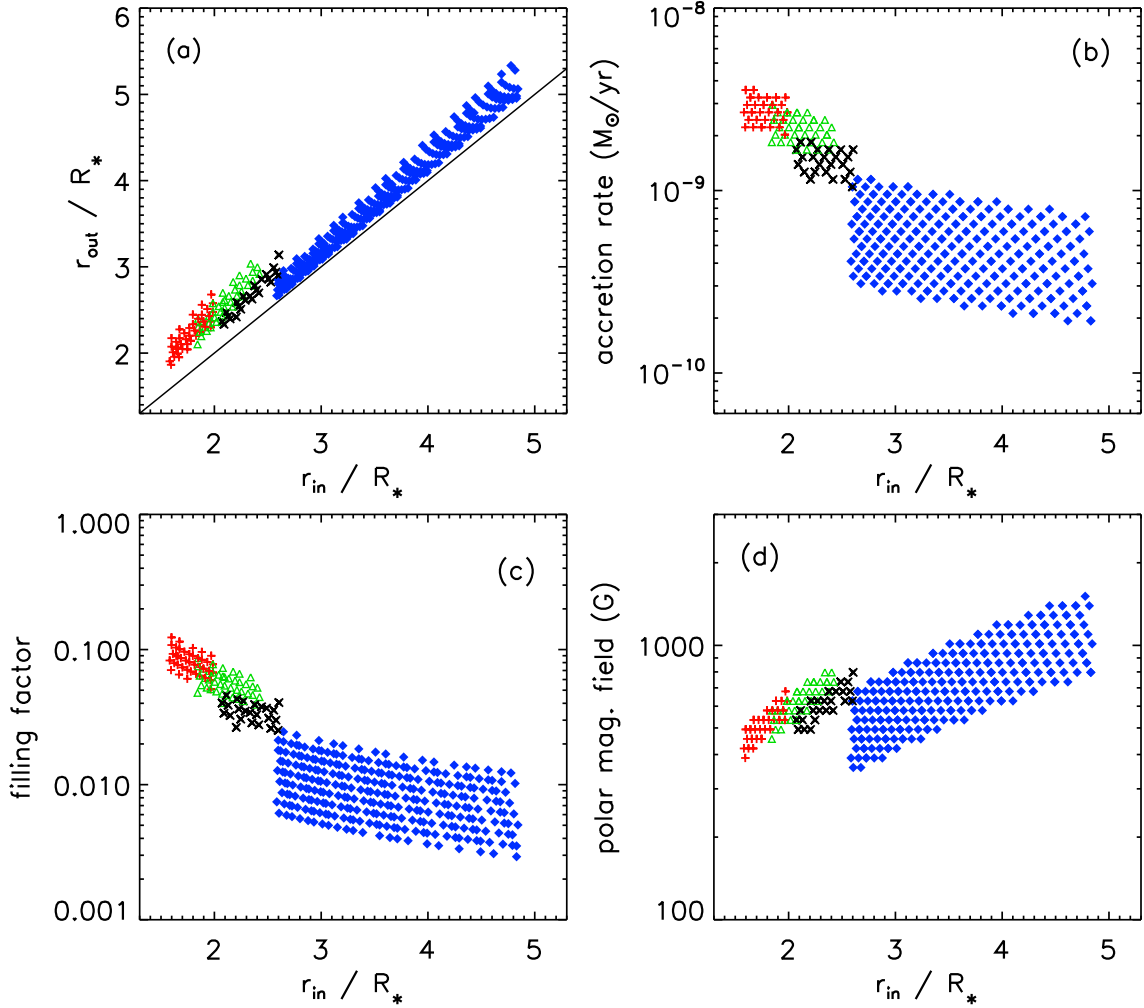}
\caption{Parameter ranges for the dipole accretion model, constrained
by \ion{Ne}{9} diagnostics from time segments 1 (JD 48.0; red crosses), 2
(JD 58.5; green triangles), 3 (JD 61.2; blue diamonds), and the summed
data (black X's). The model parameters
plotted vs the inner disk radius are outer disk radius (a), mass
accretion rate (b), filling factor at the stellar surface (c), and
polar magnetic field at the star (d). 
}
\end{figure}

\end{document}